\documentclass[english,american,aps,pre,showpacs,superscriptaddress,groupedaddress]{revtex4-1}
\usepackage[T1]{fontenc}
\usepackage[latin9]{inputenc}
\usepackage{color}
\usepackage{amsmath}
\usepackage{graphicx}
\usepackage{amssymb}
\usepackage{esint}

\makeatletter

\providecommand{\tabularnewline}{\\}

\@ifundefined{textcolor}{}
{%
 \definecolor{BLACK}{gray}{0}
 \definecolor{WHITE}{gray}{1}
 \definecolor{RED}{rgb}{1,0,0}
 \definecolor{GREEN}{rgb}{0,1,0}
 \definecolor{BLUE}{rgb}{0,0,1}
 \definecolor{CYAN}{cmyk}{1,0,0,0}
 \definecolor{MAGENTA}{cmyk}{0,1,0,0}
 \definecolor{YELLOW}{cmyk}{0,0,1,0}
 }

\makeatother

\usepackage{babel}

\begin{document}

\title{An exact formalism to study the thermodynamic properties of hard-sphere
systems under spherical confinement}

\author{Ignacio Urrutia\ref{CNEA-MC}}

\affiliation{\label{CNEA-MC}Consejo Nacional de Investigaciones Científicas y
Técnicas, Argentina (CONICET) and Departamento de Física, Comisión
Nacional de Energía Atómica, Av. Gral. Paz 1499 (RA-1650) San Martín,
Buenos Aires, Argentina.}

\email{iurrutia@cnea.gov.ar}

\author{Gabriela Castelletti\ref{IAFE}}

\affiliation{\label{IAFE}Instituto de Astronomía y Física del Espacio (CONICET-UBA),
Ciudad universitaria, Cdad. Aut. de Buenos Aires, Argentina.}
\begin{abstract}
This paper presents a modified grand canonical ensemble which provides
a new simple and efficient scheme to study few-body fluid-like inhomogeneous
systems under confinement. The new formalism is implemented to investigate
the \emph{exact} thermodynamic properties of a hard sphere (HS) fluid-like
system with up to three particles confined in a spherical cavity.
In addition, the partition function of this system was used to analyze
the surface thermodynamic properties of the many-HS system and to
derive the \emph{exact} curvature dependence of both the surface tension
and adsorption in powers of the density. The expressions for the surface
tension and the adsorption were also obtained for the many-HS system
outside of a fixed hard spherical object. We used these results to
derive the dependence of the fluid-substrate Tolman length up to first
order in density.
\end{abstract}
\maketitle

\section{Introduction}

The thermodynamic and statistical mechanical properties of a fluid
composed of many particles confined in small pores have been extensively
investigated from both a theoretical and experimental point of view
\citep{Dunkel_2006,Neimark_2006,Perret_2010}. In these systems the
fluid fills the cavity inhomogeneously and under extreme conditions,
the confinement is responsible for the dimensional crossover of the
particles \citep{Gonzalez_2006}. On the contrary, the knowledge
about the statistical mechanical properties of fluid-like systems
composed of few bodies is rather poor. Notoriously, few-body systems
have played a key role in the development of various fields of physical
science including classical, relativistic and quantum mechanics. Of
course, the fundamental advantage of the few-body systems over the
many-body ones is that only the former are (in most cases) analytically
tractable. In this sense we claim that, the importance of studying
few-body systems within the framework of the statistical mechanics
of inhomogeneous fluids relies in the fact that it provides exact
solutions in a subject where exact results are quite scarce. We find
this fact remarkable because these solutions provide new insights
to better understand the properties of many-body systems.

The grand canonical ensemble (GCE), in which the volume, temperature
and chemical potential are fixed,  is certainly the most used scheme
in the formulation of theories of fluids. In particular, the density
functional theory (DFT) formulated in the GCE, is one of the most
successful approaches used to describe the properties of inhomogeneous
fluids. Indeed, DFT calculations make it possible to analyze the adsorption
of fluids on substrates and in porous matrices as well as study the
interfacial phenomena. Other theories as the scaled particle theory
(SPT) are also formulated in the GCE. In this context hard spheres
(HS) systems play a distinguished role in developing of both, DFT
and SPT approaches, and perturbation theories of fluids \citep{Hansen2006}.
Moreover, HS systems are of particular interest as they constitute
a simplified model for both simple fluids and colloidal particles
\citep{Roth_2010}.

In this work we present a novel formulation that makes it possible
the study of few-body inhomogeneous fluids using a GCE scheme in which
the maximum number of particles is considered as a parameter of the
system. We employ the proposed GCE to perform an exact study of the
HS system with at most three particles confined in a spherical cavity.
In addition, on the basis of the new results, we analyze the surface
thermodynamic properties of the many-HS system spherically confined.
We present our result for the \emph{exact} curvature dependence of
surface (or boundary) tension and adsorption, obtained in powers of
the density. The formulation proposed here is found to be a valuable
tool for describing highly confined fluids constituted by few bodies.
Moreover, it will potentially serve to improve the DFT and SPT descriptions
associated with the surface-related thermodynamic properties and
their curvature dependence.

\section{Partition function of the few-body open system\label{sec:Part-func}}

Consider a one-component fluid containing a non-fixed number of particles
which evolve inside a region of the space that we will refer to as
$\mathcal{B}$. Such a region is determined by an external potential
$\varphi(\mathbf{r})$ that takes finite values in $\mathcal{B}$
and diverges outside it. The boundary of $\mathcal{B}$ is assumed
to be at a constant and uniform temperature $T$. In these circumstances
we expect the fluid will attain an equilibrium state in which its
physical properties either remain constant or fluctuate around a fixed
value. Furthermore, for an inhomogeneous fluid system in an equilibrium
state we make the following usual assumptions: i) if the exact expression
for the grand canonical partition function (GCPF) $\Xi$ is known,
then the exact grand potential function is given by\begin{equation}
\Omega=-\beta^{-1}\ln\Xi\:,\label{eq:grndpot}\end{equation}
where $\beta=\left(k_{B}T\right)^{-1}$ is the inverse temperature
(being $k_{B}$ and $T$ the Boltzmann's constant and the temperature,
respectively); ii) a given thermodynamic property of the system, namely
$X$, can be identified with its mean (Gibbsian) ensemble value $\left\langle X\right\rangle $.
Additionally, it is expected that $\left\langle X\right\rangle $
to be equal to the time averaged value of $X$ over an interval $\tau$,
$\bar{X}_{\tau}=\frac{1}{\tau}\int_{\tau}X(t)dt$ (at least for a
value of $\tau$ large enough). Typically, $X$ may be the energy
$E$, the pressure $p$, the number of particles $N$, etc. Standard
statistical mechanical demonstrations show that the derivatives of
the grand potential are related to the mean ensemble values $\left\langle X\right\rangle $
of most thermodynamic quantities. Other thermodynamic quantities may
be computed from the grand potential and its derivatives by applying
usual thermodynamic relations \citep{Hill1956,Hansen2006}.

In order to study few-body inhomogeneous systems, we now consider
a container $\mathcal{B}$ with at most $\mathcal{M}$ particles,
which in what follows will be a fixed parameter. The GCPF of this
system is given by\begin{equation}
\Xi_{\mathcal{M}}=\sum_{j=0}^{\mathcal{M}}\mathtt{I}_{j}z^{j}Z_{j}\:,\label{eq:GPF}\end{equation}
where $\mathcal{M}$ is the cutoff value for the maximum number of
particles. The indistinguishability factor $\mathtt{I}_{j}$ is either
equal to $\mathtt{I}_{j}=1/j!$ for indistinguishable particles or
$\mathtt{I}_{j}=1$ for distinguishable particles. In addition, $z=\Lambda^{-3}\exp\beta\mu$
is the activity being \foreignlanguage{english}{$\Lambda=h/(2\pi m\, k_{B}T)^{1/2}$}
the thermal de Broglie wavelength, while $\mu$, $m$ and $h$ are
the chemical potential, the mass of each particle and the Planck's
constant, respectively. Finally, the configuration integral (CI) of
a system with $j$ particles is\begin{equation}
Z_{j}=\int\ldots\int\,\prod e_{l}\prod e_{lm}\, d^{j}\mathbf{r}\:,\label{eq:Zjinteg}\end{equation}
with $Z_{0}=1$. Here, $e_{l}=\exp\left[-\beta\varphi(\mathbf{r}_{l})\right]$
and $e_{lm}=\exp\left[-\beta\phi(\mathbf{r}_{lm})\right]$ are the
Boltzmann's factors corresponding to the external and pair interactions,
respectively. Note that, the integration domain in Eq. (\ref{eq:Zjinteg})
is the complete space due to the fact that the spatial confinement
of the particles in the region $\mathcal{B}$ is considered in the
$e_{l}$ terms.

Because of $\Xi_{\mathcal{M}}$ is the GCPF of our system, in Eq.
(\ref{eq:grndpot}) we can replace $\Xi$ with $\Xi_{\mathcal{M}}$
to define $\Omega$. The knowledge of $Z_{j}$ for $1\leq j\leq\mathcal{M}$
along with Eq. (\ref{eq:grndpot}) and Eq. (\ref{eq:GPF}) provide
the exact statistical-thermodynamic properties of the system. In Eq.
(\ref{eq:GPF}), the expression for $\Xi_{\mathcal{M}}$ is an incomplete
(or restricted) version of the usual GCPF obtained from $\Xi=\underset{\mathcal{M}\rightarrow\infty}{\lim}\Xi_{\mathcal{M}}$.
A different restricted GCE was introduced by Yang and Lee \citep{Yang_1952}
in their study of the condensation of gases. It was used later by
Woods et al. \citep{Woods_1988} to study the adsorption of fluids
in cavities. We emphasize that in their analysis $\mathcal{M}$ represents
the maximum number of particles that fit in $\mathcal{B}$, instead
of an externally imposed parameter as we have assumed in our formalism.
An interesting property of $\Xi_{\mathcal{M}}$, Eq. (\ref{eq:GPF}),
is that for a cavity $\mathcal{B}$ with a fixed size and temperature
we obtain $\underset{z\rightarrow\infty}{\lim}\Xi_{\mathcal{M}}\propto\mathtt{I}_{j}z^{j}Z_{j}$
where $j$ denotes the maximum number of particles that hold in $\mathcal{B}$
(under the constraint $1\leq j\leq\mathcal{M}$).

\section{The few HS system in a spherical confinement}

Few-body fluid-like systems of HS with one or two particles confined
in cavities with different geometries, including the spherical, cuboidal
and cylindrical cases, were analytically solved in the canonical ensemble
in Refs. \citep{Urrutia_2008,Urrutia_2010,Urrutia_2010b}. Those works
make it possible to explore on exact grounds the corresponding open
systems with $\mathcal{M}=2$. Further, the three-body HS system was
exactly solved for the case of a spherical confinement \citep{Urrutia_2011_b,Urrutia_2011_be}.
In the current work we focus on the study of the HS open system with
$\mathcal{M}=3$ in a spherical cavity. Inhomogeneous systems with
a spherical geometry are ubiquitously present in nature in the form
of bubbles and drops. One remarkable characteristic of this symmetry
is that it does not place any preferential direction on the space.
In spite of the relevance of the spherical geometry several fundamental
questions are still unresolved regarding the curvature dependence
of the thermodynamic properties in spherical systems. In view of this,
much efforts have been directed in recent years to address this fundamental
issue.

\noindent In our description we consider an external hard-wall potential,
which is null in a spherical region $\mathcal{B}$ and diverges outside
it. The Boltzmann's factor is $e_{l}=\Theta(R-r_{l})$, where $\Theta(r)$
is the Heaviside function, $r_{l}$ is the distance from the center
of the pore to the particle $l$, and $R$ is the effective radius
of the pore \citep{Urrutia_2011_b}. The volume of $\mathcal{B}$
is $V=Z_{1}=\frac{4\pi}{3}R^{3}$ while $A=4\pi R^{2}$ is the surface
area of its boundary. By using the expressions for $Z_{2}$ and $Z_{3}$
presented in Refs. \citep{Urrutia_2008,Urrutia_2011_b,Urrutia_2011_be}
and assuming indistinguishable particles, i.e., $\mathtt{I}_{j}=1/j!$,
we obtain the exact expression for $\Xi_{3}$\begin{equation}
\Xi_{3}=1+Z_{1}z+Z_{2}\frac{z^{2}}{2}+Z_{3}\frac{z^{3}}{6}\:.\label{eq:GPF3}\end{equation}
After fixing the diameter of the particles to be $\sigma=1$ to lighten
our notation, the expressions for $Z_{2}$ and $Z_{3}$ are given
by\begin{equation}
Z_{2}=\begin{cases}
\left(\frac{2\pi}{3}\right)^{2}\left(R-\tfrac{1}{2}\right)^{3}\left(1+6R+6R^{2}+4R^{3}\right)\,, & \textrm{if }R\geq\tfrac{1}{2}\\
0\,, & \textrm{if }0\leq R<\tfrac{1}{2}\,,\end{cases}\label{eq:Z2F}\end{equation}
\begin{equation}
Z_{3}=\begin{cases}
h-\Delta\tau\,, & \textrm{if }R\geq1\\
h\,, & \textrm{if }\tfrac{1}{\sqrt{3}}\leq R<1\\
0\,, & \textrm{if }0\leq R<\tfrac{1}{\sqrt{3}}\,,\end{cases}\label{eq:Z3F}\end{equation}
with\begin{eqnarray}
h & = & \frac{\pi^{2}}{70}\left[\frac{1}{9}q\left(65+183R^{2}-342R^{4}+240R^{6}\right)-\frac{9}{2}p_{1}\left(5+12R^{2}\right)\right.\nonumber \\
 &  & +p_{2}R\left(105-280R^{2}+840R^{4}-1152R^{6}+640R^{8}\right)\Bigr]\:,\label{eq:h}\end{eqnarray}
\begin{equation}
q=\sqrt{3R^{2}-1}\:,\quad p_{1}=\arctan\left(2q\right)\:,\quad p_{2}=\arctan\left(q/R\right)\:,\label{eq:qp2p3}\end{equation}
\begin{equation}
\Delta\tau=\frac{128}{189}\pi^{3}\left(R-1\right)^{5}\left(1+5R+\frac{69}{10}R^{2}+5R^{3}+R^{4}\right)\:.\label{eq:DTau3R}\end{equation}
Concerning the basic geometrical properties of the spherical confinement
we find that, a cavity with $R<\frac{1}{2}$ can contain at most one
particle, while for the cases $R<\frac{1}{\sqrt{3}}\doteq0.57735$,
$R<\sqrt{\frac{3}{8}}\doteq0.612372$ and $R<\frac{1}{\sqrt{2}}\doteq0.707107$,
a maximum of two, three and four particles can be fitted in the cavity,
respectively. These values imply that by evaluating $\Xi_{3}$ we
have not only solved the restricted system with $\mathcal{M}=3$ and
any value of $R$, but also we have obtained the exact solution of
the unrestricted system for $R<\sqrt{\frac{3}{8}}$ because in this
case $\Xi(R)=\Xi_{3}(R)$.

The thermodynamic fundamental relations for the grand potential are\begin{equation}
\Omega=U-TS-\mu N\:,\label{eq:Omega}\end{equation}
\begin{equation}
d\Omega=-S\, dT-P_{W}A\, dR-N\, d\mu\:,\label{eq:dOmega}\end{equation}
here $S$ is the entropy, $dV$ is equal to $A\, dR$, $P_{W}$ represents
the (mean) pressure on the wall also called the work-pressure, and
$dW=P_{W}dV$ is the total reversible work performed by the system
on its environment. Given that the system is athermal, it is preferable
to use temperature-independent quantities like $\beta\Omega$ to conduct
the analysis of its properties. Therefore, some of the usual thermodynamic
and statistical mechanical relations for systems under spherical confinement
can be written as\begin{equation}
\beta P_{W}=-A^{-1}\left.\frac{\partial\,\beta\Omega}{\partial R}\right|_{\beta,z}\:,\label{eq:Pw}\end{equation}
\begin{equation}
N\equiv\left\langle N\right\rangle =-z\left.\frac{\partial\,\beta\Omega}{\partial z}\right|_{\beta,R}\:,\label{eq:N}\end{equation}
\begin{equation}
\sigma_{_{N}}\!^{2}\equiv\left\langle N^{2}\right\rangle -N^{2}=z\left.\frac{\partial N}{\partial z}\right|_{\beta,R}\:.\label{eq:sgNsqr}\end{equation}
Here $\sigma_{_{N}}$ is the standard deviation in the number of particles
which quantifies the spontaneous fluctuation of $N$. Since the energy
of the system is equal to that of the classical ideal gas, i.e., $\beta U=\frac{3}{2}N$,
using Eqs. (\ref{eq:Omega}) and (\ref{eq:N}) we can also calculate
$S$. At this stage, the exact properties obtained for the system
are functions of $z$ and $R$. One important point is the fact that
the few-body HS open system is in chemical equilibrium with a bulk
HS fluid. Unfortunately, the exact properties of the bulk HS fluid
are not analytically known, and thus, to present our results in terms
of the bulk density $\rho_{b}$ (instead of $z$) we adopt the simple
and accurate Carnahan-Starling's equation of state, which for $z$
gives \citep{Carnahan_1969,Gonzalez_2006}\begin{equation}
z_{\textrm{CS}}=\rho_{b}\times\exp\left[\frac{8\eta_{b}-9\eta_{b}^{2}+3\eta_{b}^{3}}{\left(1-\eta_{b}\right)^{3}}\right]\:,\label{eq:zCS}\end{equation}
where $\eta_{b}=\frac{\pi}{6}\rho_{b}\sigma^{3}$ is the corresponding
packing fraction. %
\begin{table}[t]
\centering{}\begin{tabular}{|c|c|c|c|c|c|c|c|}
\hline 
$z$ & $0.1$ & $0.2$ & $0.5$ & $1$ & $2$ & $5$ & $10$\tabularnewline
\hline 
$\rho_{b}$ & $0.072226$ & $0.11583$ & $0.18962$ & $0.25142$ & $0.31323$ & $0.39028$ & $0.44353$\tabularnewline
\hline 
$z$ & $20$ & $50$ & $100$ & $200$ & $500$ & $10^{3}$ & $10^{5}$\tabularnewline
\hline 
$\rho_{b}$ & $0.49217$ & $0.54981$ & $0.58888$ & $0.62452$ & $0.66704$ & $0.69619$ & $0.84313$\tabularnewline
\hline
\end{tabular}\caption{Activity values $z$, used for the curves plotted in Fig. \ref{fig:Pw}
and the corresponding bulk density values, $\rho_{b}$.\label{tab:TI}}
\end{table}
We remark that Eq. (\ref{eq:zCS}) is the unique approximated relation
introduced in this section. Although it is exact only up to third
order in density accurately describes the bulk system for all the
fluid-phase density range. As a reference, we mention that in the
bulk HS system the fluid (disordered) and solid (ordered) stable phases
coexist at $\beta P\simeq11.576$, being the fluid and the solid bulk-densities
$\rho_{bf}\simeq0.943$ and $\rho_{bs}\simeq1.041$, respectively
\citep{Kolafa_2004,Bannerman_2010}.

\begin{figure}[t]
\begin{centering}
\includegraphics[width=7cm]{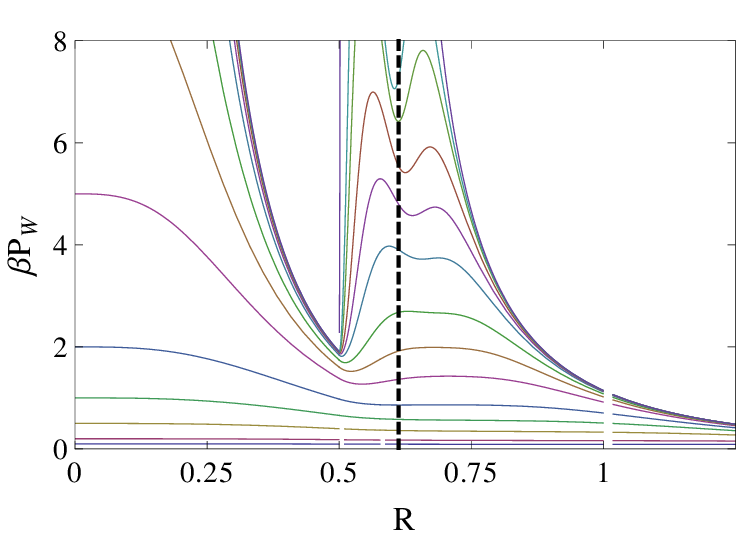} 
\par\end{centering}

\caption{Pressure on the wall against the radius of the pore obtained for values
of the activity $z$, and bulk density $\rho_{b}$ listed in Table
\ref{tab:TI} ($z$ and $\rho_{b}$ increase from bottom to top).
The dashed line indicates the radius $R=0.612$ for which a maximum
of three particles can be fitted in the cavity.\label{fig:Pw}}
\end{figure}
Fig. \ref{fig:Pw} displays the pressure on the wall as a function
of the radius of the cavity, considering different values of $\rho_{b}$.
Table \ref{tab:TI} summarizes the exact values of $z$ used in plotting
each curve together with the corresponding approximated values of
$\rho_{b}$ obtained from the numerical inversion of Eq. (\ref{eq:zCS}).
The dashed vertical line in Fig. \ref{fig:Pw} corresponds to $R=0.612$,
that is, the maximum value for which our results are exact even for
the unrestricted system. From this graphic it is evident that the
pressure on the wall has a non-monotonic behavior with loops characterized
by regions of negative and positive slopes, which are related to the
mechanical instability of the system.%
\begin{figure}[t]
\begin{centering}
\includegraphics[width=7cm]{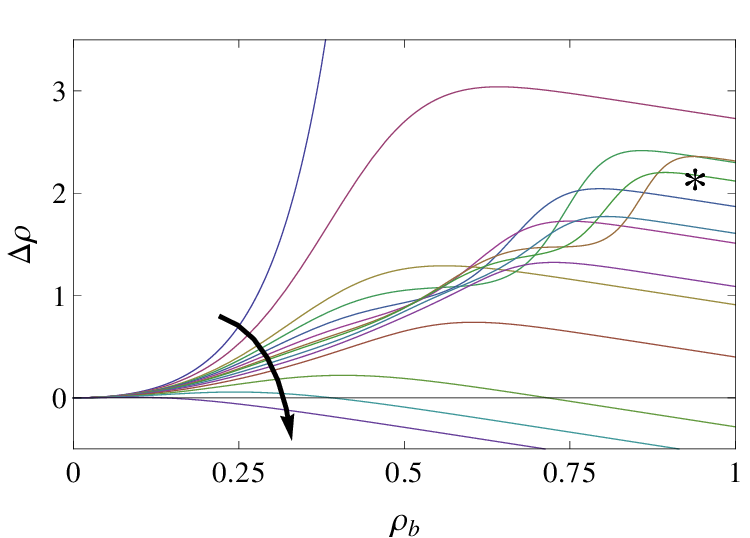}
\par\end{centering}

\caption{Relative adsorption in the pore as a function of the bulk density
plotted for different radii of the pore $R=$$0.2,0.4,0.5,$ $0.525,0.55,0.575,$
$0.6,0.61237,$ $0.65,0.7,0.8,$ $1,1.2,$ and $1.5$. The arrow marks
the direction of increasing values of $R$. The asterisk indicates
the curve for $R=0.61237$.\label{fig:deltarho}}
\end{figure}
 In Fig. \ref{fig:deltarho} we plot the relative adsorption in the
pore per unit volume as a function of $\rho_{b}$, that is $\Delta\rho=\hat{\rho}-\rho_{b}$
where $\hat{\rho}=N/V$. There, the arrow indicates increasing values
of $R$. The negative adsorption in the case of the curves corresponding
to $R=$$1,1.2,$ and $1.5$ is a consequence of the cutoff in the
maximum number of particles contained in the pore. This also applies
to regions with positive adsorption and negative slope observed in
this plot for $R=$$0.65,0.7,0.8,$ and $1$.%
\begin{figure}[t]
\centering{}\includegraphics[width=7cm]{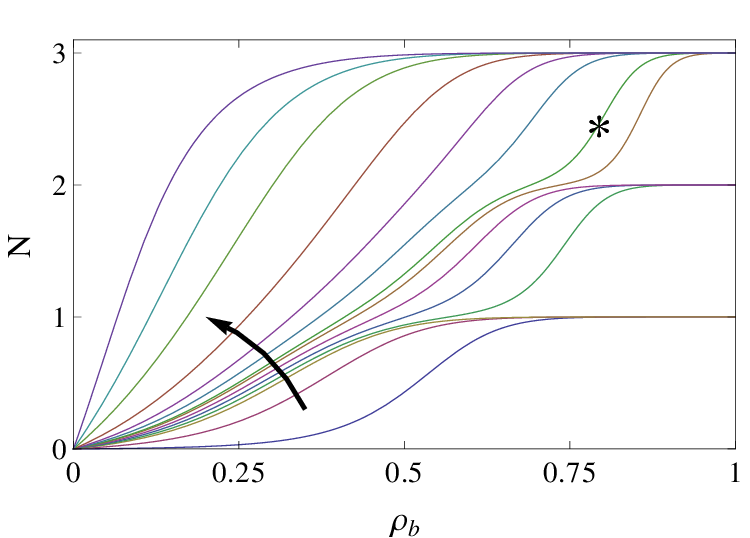}\caption{Mean number of particles vs. bulk density for different radii of the
pore. The arrow, the asterisk and the values of $R$ were described
in Fig. \ref{fig:deltarho}.\label{fig:N}}
\end{figure}
\begin{figure}[t]
\begin{centering}
\includegraphics[width=7cm]{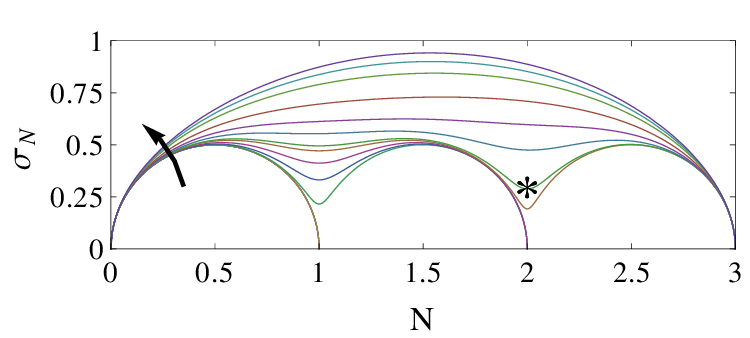}
\par\end{centering}

\caption{Standard deviation in $N$ as a function of $N$ for different radii
of the pore. The arrow, the asterisk and the values of $R$ were described
in Fig. \ref{fig:deltarho}.\label{fig:circles}}
\end{figure}
Fig. \ref{fig:N} shows the filling of the cavity as a function of
$\rho_{b}$ for different radii of the pore. For each value of $R$
considered the saturation value of $N$ corresponds to the maximum
number of particles that holds in the cavity (at most three particles).
As it can be observed, $N(\rho_{b},R)$ is a non-decreasing function.
It is noticeable that for $0.5<R<0.61237$ the curves change their
concavity several times with the increase of $\rho_{b}$. This property
determines the behavior of $\sigma_{N}^{2}(\rho_{b})$ through Eq.
(\ref{eq:sgNsqr}). Fig. \ref{fig:circles} illustrates the relation
between the mean number of particles in the cavity and its fluctuation
$\sigma_{_{N}}$. Each curve corresponds to the parametric plot $(N(z),\sigma_{N}(z))$
for a fixed value of $R$. We find that all the curves corresponding
to $R<0.5$ collapse onto the semi-circumference centered at $N=0.5$.
For $0.5<R<0.577$ and N increasing in value $1.5<N\leq2$ the curves
approach to the second semi-circumference centered at $N=1.5$. In
the range $2.5<N\leq3$ a third semi-circumference centered at $N=2.5$
is fitted by the curves corresponding to $R>0.577$. In addition,
we have analyzed the case $\mathcal{M}=2$ and obtained a picture
similar to that displayed for the case $\mathcal{M}=3$ but with only
two semi-circumferences centered at $N=0.5$ and $N=1.5$ (the graphic
is not shown here). One important consequence of the graphic shown
in Fig. \ref{fig:circles} is that it can be straightforwardly extended
to obtain the representation for the case with a value $\mathcal{M}>3$.
Basically, it will involve a sequence of $\mathcal{M}$ semi-circumferences
of unit diameter centered at half-integer values of $N$ along the
abscissa axis. In particular, for a given $R$ the curve will start
at the origin $(N,\sigma_{N})=(0,0)$ and will collapse either onto
the semi-circumference that ends at the value $N$ equal to the maximum
number of particles that fit in the cavity or onto the $\mathcal{M}$-th
semi-circumference in case $R$ is large enough.

It is to be noted that the outmost curve which connects $N=0$ with
$N=\mathcal{M}$ can be analytically obtained by considering the limit
of a very large cavity, $V\rightarrow\infty$ (infinite dilution).
In this low density limit we can replace $Z_{i}$ with $Z_{1}^{i}=V^{i}=\left(\frac{4\pi}{3}R^{3}\right)^{i}$
and hence $N$ and $\sigma_{N}$ become functions only of the parameter
$a=z\, Z_{1}$. By following this procedure we effectively reduce
the system to an ideal gas (with up to $\mathcal{M}$ particles).
The exact analytic expressions for $\sigma_{N}(a)$ and $N(a)$ with
$\mathcal{M}=3$ are parametrically plotted (using $a$ as a parameter)
in the outmost curve in Figure 4. As a closing remark, this analysis
shows that \emph{any} system with at most one particle is described
by the first semi-circumference, irrespective of the nature of the
particle and the confinement.

\section{The many HS system onto a spherical wall}

In what follows we draw attention to the implications that the expression
for $\Omega$, obtained for the system with $\mathcal{M}=3$, has
on the surface-thermodynamic properties of the many-body system. In
particular, we will address on the HS system in contact with a hard
spherical wall in the low density limit. For each magnitude of the
system, our main goal is to find its \emph{exact} power series in
$\rho_{b}$ up to the three-body term, with the dependence on the
curvature included through the coefficients of the series. Our procedure
is based on the ideas developed in Ref. \citep{Urrutia_2010b}, which
enable us to express the analytically known terms $Z_{i}(R)$ as a
function of the set of measures $\{V,A,R\}$. We begin by placing
the dividing surface at $R$ and we use the same decomposition rule
adopted in Ref. \citep{Urrutia_2011_b} (Eqs. (28) and (29) therein).
For the unknown $Z_{i}$ terms we assume a generic dependence $Z_{i}(V,A,R)$.
As a result, we obtain an expression for $\beta\Omega$ that is a
function of $V,A,R$ and $z$. We then fix $\mathcal{M}=4$ in Eq.
(\ref{eq:GPF}) and calculate the magnitude $X$. By following standard
procedures based on the inversion and composition of the power series
we expand the expression for $X$ as a power series in the variable
$\rho_{b}$. Then, we truncate the series to the first term in which
$Z_{4}$ appears. The exact series of $z(\rho_{b})$ truncated to
third order in density, $z=\rho_{b}+\frac{4\pi}{3}\rho_{b}^{2}+\frac{47\pi^{2}}{36}\rho_{b}^{3}$,
is sufficient for our purposes as we have verified that the inclusion
of the fourth order term does not modify our results. In addition,
to obtain the planar-surface property we consider in the series the
limit $R\rightarrow\infty$. Thus, we determine in the planar limit
the expressions for the fluid-substrate surface tension $\gamma\equiv\partial\Omega/\partial A|_{V,R}$
and the adsorption per unit area $\Gamma\equiv V\left(\hat{\rho}-\rho_{b}\right)/A$,

\begin{equation}
\beta\gamma_{\infty}=-\frac{\pi}{8}\rho_{b}^{2}\left(1+\frac{149\pi}{210}\rho_{b}\right)+O(\rho_{b}^{4})\:,\label{eq:gamma}\end{equation}
\begin{equation}
\Gamma_{\infty}=\frac{\pi}{4}\rho_{b}^{2}\left(1-\frac{113\pi}{420}\rho_{b}\right)+O(\rho_{b}^{4})\:.\label{eq:Gamma}\end{equation}

\noindent Eqs. (\ref{eq:gamma}) and (\ref{eq:Gamma}) exactly reproduce
the first two terms (the only known up to present) of the series expansion
of $\beta\gamma_{\infty}(\rho_{b})$ and $\Gamma_{\infty}(\rho_{b})$
\citep{Bellemans_1962,Siderius_2006} and are consistent with the
Gibbs adsorption isotherm $\Gamma_{\infty}=-\partial\gamma_{\infty}/\partial\mu$
to order $\rho_{b}^{3}$.

\begin{figure}[t]
\noindent \centering{}\includegraphics[width=7cm]{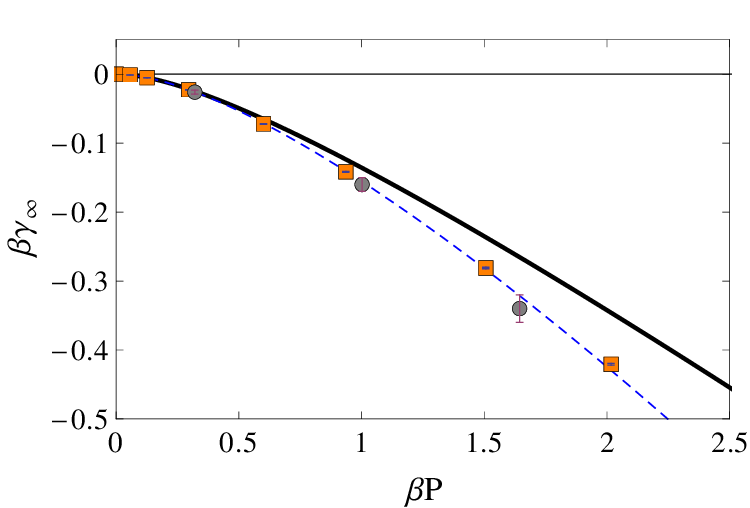}\caption{Fluid-substrate surface tension vs. pressure. The filled circles and
squares correspond to simulation and thermodynamic integration data
from \citep{Heni_2001} and \citep{Laird_2010}, respectively. The
dashed line shows SPT values \citep{Siderius_2007_b}. The parametric
plot of the power series $\beta P(\rho_{b})$ and $\beta\gamma_{\infty}(\rho_{b})$
is shown by the continuous line.\textcolor{red}{\label{fig:Surface-Tension}}}
\end{figure}

\noindent In Fig. \ref{fig:Surface-Tension} we show a plot of the
surface tension vs. pressure in the low pressure region, which is
also the low-density region. Our results are represented parametrically
$(\beta\gamma_{\infty}(\rho_{b}),\beta P(\rho_{b}))$ using both,
Eq. (\ref{eq:gamma}) and the exact third order virial series $\beta P=\rho_{b}+\frac{2\pi}{3}\rho_{b}^{2}+\frac{5\pi^{2}}{18}\rho_{b}^{3}$.
The dashed line included in this figure plots previous SPT results
taken from Fig. 2 in Ref. \citep{Siderius_2007_b}. Since all the
plotted SPT versions are nearly indistinguishable in the range of
pressure considered, we only include the results of the CS-SPT and
CS-SPT$_{M}$ (they collapse on a unique curve) to make easy the examination
of consistency between SPT and our result. Also, the simulation and
thermodynamic integration data from \citep{Heni_2001,Laird_2010}
are indicated by filled circles and squares. We note that simulation
results fall nicely on the SPT curve. On the other hand, the exact
low-density series adequately fits numerical results over the range
$0<\beta P\lesssim0.6$.%

\noindent As it is well known, the relevant problem in statistical
mechanics regarding the dependence of the surface thermodynamic properties
on the curvature remains unresolved, even for the HS fluid in contact
with hard spherical walls. However, following the procedure described
above and writing the magnitude $X/X_{\infty}$ as a double power
series in the variables $R^{-1}$ and $\rho_{b}$, we obtain\begin{equation}
\gamma(R)/\gamma_{\infty}=1+\delta_{\infty}\,2R^{-1}+\delta_{\mathsf{k}}R^{-2}+O(R^{-3})\:,\label{eq:gamR}\end{equation}
\begin{equation}
\Gamma(R)/\Gamma_{\infty}=1+\xi_{\mathsf{j}}\,2R^{-1}+\xi_{\mathsf{k}}R^{-2}+O(R^{-3})\:.\label{eq:GamR}\end{equation}
The complete expressions for $\gamma(R)$ and $\Gamma(R)$ in the
limit of both small $\rho_{b}$ and large $R$ is presented in Appendix
\ref{sec:AppendixA}. In the latter two equations, $2R^{-1}$ and
$R^{-2}$ are the mean and Gaussian curvatures of the spherical cavity,
respectively. The coefficient $\delta_{\infty}$ denotes the radius-independent
wall-fluid Tolman length. In fact, neither of the coefficients in
these equations depends on $R$. We verified that Eq. (\ref{eq:gamR})
and Eq. (\ref{eq:GamR}) hold for fluids in contact with a hard spherical
wall, irrespective of whether the fluid is inside the cavity ($R>0$)
or outside of a fixed hard sphere ($R<0$, also known as an empty
cavity in a bulk fluid) \citep{Urrutia_2011_b}. To first order in
density the coefficients $\delta_{\infty}$ and $\delta_{k}$ can
be computed from\begin{eqnarray}
\delta_{\infty}\doteq0.34299\,\rho_{b}\,, &  & \delta_{\mathsf{k}}\doteq-0.05555+0.19933\,\rho_{b}\,,\nonumber \\
\xi_{\mathsf{j}}\doteq0.51448\,\rho_{b}\,, &  & \xi_{\mathsf{k}}\doteq-0.05555+0.29899\,\rho_{b}\,.\label{eq:dTol}\end{eqnarray}
No numerical simulation results have been published up to present
for these magnitudes. The expressions for both $\delta_{\infty}$
and $\delta_{\mathsf{k}}$ determined by the above procedure are consistent
with those recently reported in \citep{Urrutia_2011_b}, derived from
an exact calculation for a few-body system treated within the framework
of the canonical ensemble.%
\begin{figure}[t]
\noindent \centering{}\includegraphics[width=7cm]{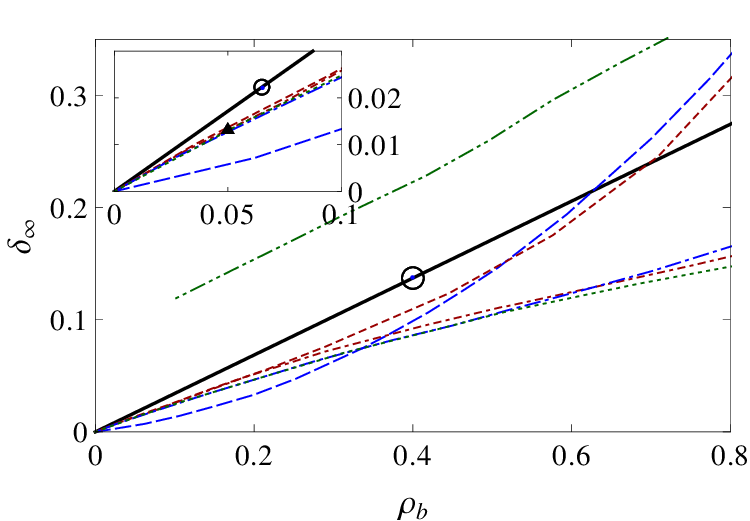}\caption{Tolman length vs. density. The continuous curve with the open circle
corresponds to the linear dependence in $\rho_{b}$ from this work.
The six SPT versions from Refs. \citep{Chatterjee_2006,Siderius_2007_b}
are shown by non-continuous curves: SPT$_{3}$ (dotted line), SPT$_{6}$
(short dashed line), CS-SPT (dashed line), SPT$_{6M}$ (short dashed-dot
line), CS-SPT$_{M}$ (dashed-dot line), and ESPT (dashed-dot-dot line
that lies above all other lines). The graphic includes at the top
left corner a close up of the curves around $\rho_{b}=0$.\label{fig:Tolman-length.}}
\end{figure}
 Figure \ref{fig:Tolman-length.} displays the Tolman length vs. density.
There, the continuous line with the open circle shows the power series
of $\delta_{\infty}(\rho_{b})$ around $\rho_{b}=0$ truncated to
the linear order, Eq. (\ref{eq:dTol}). This figure also includes
the estimates of $\delta_{\infty}$ obtained by Siderius et al. \citep{Siderius_2007_b}
and Chatterjee et al. \citep{Chatterjee_2006} using six different
versions of SPT, which are plotted using different styles of non-continuous
lines. The first observation we make is that the general trend of
the SPT curves is well reproduced by our result. In what follows
we will concentrate in the low density region of this figure for which
we know the exact values of the $\delta_{\infty}$-intercept and the
slope at $\rho_{b}=0$ of the \emph{true} $\delta_{\infty}(\rho_{b})$
function. By adopting the nomenclature used in Ref. \citep{Siderius_2007_b},
we note that the SPT proposals CS-SPT$_{M}$, SPT$_{6}$, SPT$_{6M}$
and SPT$_{3}$ are nearly indistinguishable as $\rho_{b}$ becomes
$0$. In view of this we address our analysis of these four curves
directly to the study of CS-SPT$_{M}$. Interestingly, among the six
SPT theories, the CS-SPT and CS-SPT$_{M}$ are the unique ones that
provide analytic formulae for $\delta_{\infty}(\rho_{b})$, and thus
we can determine the corresponding series expansion. For CS-SPT and
CS-SPT$_{M}$ we obtain slopes of $0.0970$ and $0.2675$, respectively.
The comparison of CS-SPT and its modified version CS-SPT$_{M}$ with
the exact result, $0.34299$, reveals that the modified approach to
SPT significantly improves the description of the $\delta_{\infty}$
behavior at low density. Regarding the ESPT curve for $\delta_{\infty}(\rho_{b})$,
its behavior is unknown in the range $0<\rho_{b}<0.1$.

\section{Summary}

In this work we have introduced a new formalism to analyze the statistical
mechanics of few-body open systems. It is based on a grand canonical
ensemble with a restriction in the maximum number of particles. The
main feature of the new GCE formalism is that it incorporates the
fluctuation in the number of particles, measured by $\sigma_{N}$.
On the basis of this new grand canonical scheme we have calculated
the partition function of the HS fluid-like system composed of at
most three particles confined in a spherical cavity. Using this partition
function we obtained the analytic expressions for several thermodynamic
properties of the system. Among all the properties analyzed for this
system, the dependence of $\sigma_{N}$ on $N$ is, in virtue of its
simplicity, symmetry and beauty, the most interesting one. We are
convinced that it should deserve more attention in future studies
about the statistical mechanics of confined fluids.

We have also investigated the thermodynamic properties of the many-body
system of HS in contact with a hard spherical wall. By adopting a
power series representation we reported new expressions for both the
surface tension and adsorption, as a function of the density and curvature.
In the planar wall limit $R\rightarrow\infty$ our results are consistent
with those from previous results published in the literature. We used
the new expression for the surface tension to obtain the fluid-wall
Tolman length up to the first order in density. We conclude that the
general trend of $\delta_{\infty}$ derived from the new formalism
adequately agrees with SPT results presented by Siderius et. al. \citep{Siderius_2007_b}.
In particular, from the comparison of our exact result with that obtained
within the framework of SPT, we found that the highest level of consistency
in $\delta_{\infty}(\rho_{b})$ at zero density occurs for the CS-SPT$_{M}$,
SPT$_{6}$, SPT$_{6M}$, and SPT$_{3}$ versions although the slope
obtained in these cases differs by about $20\%$ from the exact value.
We feel that our findings concerning the description of HS systems
are important as they represent new analytic results that contribute
to better understand the long-standing problem of the thermodynamics
of confined HS systems and their surface-related properties. Moreover,
our results may provide new insights for future improvements on DFT
which are based on a reference fluid of HS.
\begin{acknowledgments}
I.U. and G.C. acknowledge the comments from the anonymous referees.
This work was supported by Argentina Grants CONICET-PIP 0546, UBACyT
20020100200156 and ANPCyT-PICT 2008-0795.
\end{acknowledgments}
\appendix

\section{\label{sec:AppendixA}}

We present here the expressions for $\beta\gamma$ and $\Gamma$ up
to the third order in density \begin{equation}
\beta\gamma=\left(a_{2}-\frac{c_{2k}}{R^{2}}\right)\text{\ensuremath{\rho_{b}}}^{2}+\left(a_{3}+\frac{8}{3}\pi a_{2}-\frac{2c_{3j}}{R}-\frac{3c_{3k}+8\pi c_{2k}}{3R^{2}}\right)\text{\ensuremath{\rho_{b}}}^{3}\:,\end{equation}
\begin{equation}
\Gamma=\left(-a_{2}+\frac{c_{2k}}{R^{2}}\right)2\text{\ensuremath{\rho_{b}}}^{2}+\left(-\frac{9a_{3}+16\pi a_{2}}{3}+\frac{8c_{3j}}{R}+\frac{9c_{3k}+16\pi c_{2k}}{3R^{2}}\right)\text{\ensuremath{\rho_{b}}}^{3}\:,\end{equation}
with $a_{2}=-\frac{\pi}{8}$, $a_{3}=\frac{137\pi^{2}}{560}$, $c_{2k}=-\frac{\pi}{2^{4}3^{2}}$,
$c_{3j}=\frac{9\pi\surd3+16\pi^{2}}{1536}$, $c_{3k}=\frac{781\pi^{2}}{36288}$.
In both equations the coefficients of $\text{\ensuremath{\rho_{b}}}^{3}$
were truncated at order $R^{-2}$.

\bibliographystyle{aipnum4-1}

%

\end{document}